\documentclass[aip, apl, reprint, 10pt, twocolumn, showpacs, preprintnumbers,superscriptaddress, amsmath, amssymb]{revtex4-2}
\usepackage[pdftex]{graphicx}
\usepackage[ngerman,english]{babel}
\usepackage{natbib}
\usepackage{color}
\usepackage{siunitx}
\usepackage{calc}
\usepackage[colorlinks=true, pdfstartview=FitV, linkcolor=blue,citecolor=blue, urlcolor=blue]{hyperref}
\usepackage{float}

\definecolor{green2}{rgb}{.0, .58, 0}

\begin{document}
\begin{abstract}
  We use a scanning superconducting quantum interference device
  (SQUID) to image the magnetic flux produced by a superconducting
  device designed for quantum computing.  The nanometer-scale
  SQUID-on-tip probe reveals the flow of superconducting current
  through the circuit as well as the locations of trapped magnetic
  flux.  In particular, maps of current flowing out of a flux-control
  line in the vicinity of a qubit show how these elements are coupled,
  providing insight on how to optimize qubit control.
\end{abstract}
\title{Magnetic imaging of superconducting qubit devices with scanning
  SQUID-on-tip}
\author{E.~Marchiori} \affiliation{Department of Physics, University
  of Basel, 4056 Basel, Switzerland} \author{L.~Ceccarelli}
\affiliation{Department of Physics, University of Basel, 4056 Basel,
  Switzerland} \author{N.~Rossi} \affiliation{Department of Physics,
  University of Basel, 4056 Basel, Switzerland} \author{G.~Romagnoli}
\affiliation{Department of Physics, University of Basel, 4056 Basel,
  Switzerland} \author{J.~Herrmann} \affiliation{Department of
  Physics, ETH Z\"urich, 8093 Z\"urich, Switzerland}
\author{J.-C.~Besse} \affiliation{Department of Physics, ETH Z\"urich,
  8093 Z\"urich, Switzerland} \author{S.~Krinner}
\affiliation{Department of Physics, ETH Z\"urich, 8093 Z\"urich,
  Switzerland} \author{A.~Wallraff} \affiliation{Department of
  Physics, ETH Z\"urich, 8093 Z\"urich, Switzerland}
\affiliation{Quantum Center, ETH Z\"urich, 8093 Z\"urich, Switzerland}
\author{M.~Poggio} \affiliation{Department of Physics, University of
  Basel, 4056 Basel, Switzerland} \affiliation{Swiss Nanoscience
  Institute, University of Basel, 4056 Basel, Switzerland}
\email{martino.poggio@unibas.ch}
\maketitle

Manufacturers of integrated circuits rely on a number of techniques to
verify complex circuit designs, locate defects, and carry out failure
analysis.  These include optical microscopy, scanning electron
microscopy, focused ion beam milling, microprobing, and thermal
imaging, which -- for example -- is used to identify the location of
short circuits via the Joule heating associated with large currents.
Among non-destructive methods, imaging magnetic fields via scanning
superconducting quantum interference device (SQUID) microscopy
combines the highest spatial resolution with the highest current
sensitivity~\cite{chatraphorn_scanning_2000,knauss_scanning_2001}.
For these reasons, such scanning probes are particularly suited for
the investigation of flux trapping and current flow in superconducting
circuits~\cite{jeffery_magnetic_1995,stan_critical_2004,jelic_imaging_2017,ceccarelli_imaging_2019}.

Here, we use a SQUID-on-tip
probe~\cite{finkler_self-aligned_2010,vasyukov_scanning_2013} to image
a superconducting circuit designed for quantum computation.  Maps of
the stray magnetic field produced by the qubit control line in the
vicinity of the transmon qubit reveal the details of the coupling, and
may provide a route towards optimization of the device.  Such maps can
be reconstructed into images of current flow, which may be useful for
the suppression of crosstalk originating from uncontrolled return
current paths, i.e.\ currents draining from a qubit control line
spuriously coupling flux to other qubits.  Quantifying and mitigating
such crosstalk is particularly relevant for the parallel execution of
flux-controlled two-qubit
gates~\cite{arute_quantum_2019,krinner_realizing_2022}.


\begin{figure}[t]
  \includegraphics[width=0.45\textwidth]{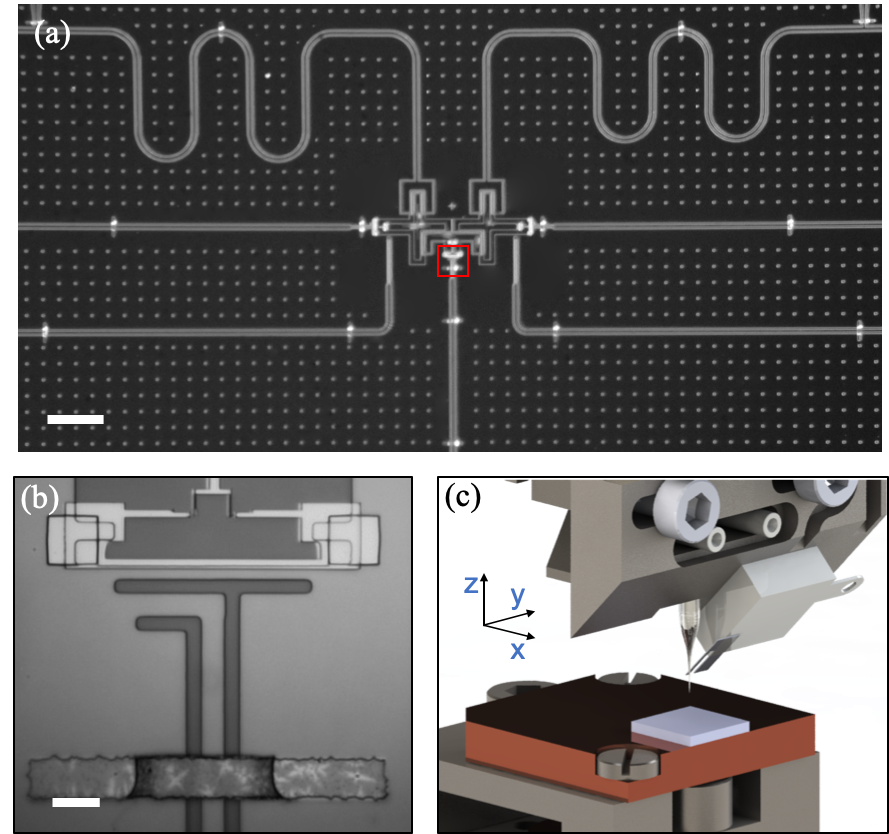}
  \caption{Superconducting qubit circuit and experimental setup.  (a)
    Optical micrograph of the circuit with the investigated area
    highlighted by the red box. Nb appears dark and exposed Si light.
    (b) Enlarged view of the investigated area, showing the base of
    the control line and the SQUID loop of the transmon qubit above
    it.  The Al SQUID loop appears lightest, the Nb ground plane
    light, and the exposed Si dark.  An Al airbridge is visible at the
    bottom.  (c) Schematic diagram of the superconducting circuit in
    the scanning SQUID microscopy
    setup~\cite{ceccarelli_scanning_2020}.  The SQUID-on-tip sensor
    measures the magnetic field along the $z$-direction, while the
    superconducting circuit is scanned underneath in the
    $xy$-plane. Scale bars: (a) \SI{200}{\micro\meter} and (b)
    \SI{10}{\micro\meter}.}
  \label{Fig1}
\end{figure}

The superconducting qubit circuit that we investigate, is patterned
from a 150-nm-thick Nb film deposited on a high-resistivity intrinsic
Si substrate.  Al airbridges are added to the device to establish a
well connected ground plane.  The circuit consists of three transmon
qubits, each with flux-control lines, where the middle qubit is used
as a tunable coupler~\cite{collodo_implementation_2020}.  The transmon
qubits and the control line to which they are coupled are shown in the
optical micrograph of Fig.~\ref{Fig1}~(a).  We focus on the roughly
$50 \times \SI{50}{\micro\meter^2}$ region at the end of the central
control line, shown in Fig.~\ref{Fig1}~(b), where the flux it
produces couples to the SQUID loop of the transmon qubit above.  The
SQUID loop of the transmon qubit is made from Al and comprises two
Josephson junctions whose electrodes are fabricated to form a closed
loop and overlap with the Nb ground plane.  In addition, electrical DC
contact between the Al SQUID electrodes and the Nb ground plane is
established using Al bandages~\cite{dunsworth_characterization_2017}.

In order to map the stray magnetic field perpendicular to the plane of
the superconducting circuit, we carry out scanning SQUID microscopy
(SSM) with a SQUID-on-tip probe mounted in a high-vacuum microscope
operating at \SI{4.2}{\kelvin}, shown schematically in
Fig.~\ref{Fig1}~(c).  The probe is fabricated by evaporating Pb on the
apex of a pulled quartz capillary according to a self-aligned method
pioneered by Finkler et al.~\cite{finkler_self-aligned_2010} and
perfected by Vasyukov et al.~\cite{vasyukov_scanning_2013} The
SQUID-on-tip used here has an effective loop-diameter of
\SI{190}{\nano\meter}, as extracted from measurements of the critical
current $I_{\text{SOT}}$ as a function of a uniform magnetic field
$\mathbf{B}_a = B_a \mathbf{\hat{z}}$, applied perpendicular to the
SQUID loop.

The superconducting qubit circuit is mounted in the SSM in a plane
parallel to and just below the loop of SQUID-on-tip, as shown in
Fig.~\ref{Fig1}~(c).  Since the current response of the SQUID-on-tip
is proportional to the magnetic flux threading through its SQUID loop,
it provides a measure of the component of the local magnetic field
perpendicular to the circuit plane $B_z$, integrated over the loop.  A
serial SQUID array amplifier (Magnicon) is used to measure the current
flowing through the SQUID-on-tip~\cite{ceccarelli_imaging_2019}.  The
sample is positioned using piezoelectric positioners and scanners
(Attocube).  By scanning the sample at constant tip-sample spacing, we
map $B_z (x, y)$ with sub-micron spatial resolution that is limited by
this spacing and -- ultimately -- by the SQUID-on-tip diameter.

In general, a map of magnetic field cannot be reconstructed into a map
of its source current density by simply inverting the Biot-Savart law,
because three-dimensional current densities do not produce unique
magnetic field patterns.  In-plane current densities or current
densities in thick films that are uniform throughout their thickness,
however, can be uniquely determined by their magnetic field.
Therefore, by assuming a uniform current density $\mathbf{J}_{xy}$
along the thickness of the superconducting Nb film, which is of the order of
its penetration depth, we can reconstruct $\mathbf{J}_{xy} (x, y)$
from the measured
$B_z(x,
y)$~\cite{roth_using_1989,chang_nanoscale_2017,marchiori_nanoscale_2022}.

\begin{figure}[t]
  \includegraphics[width=0.45\textwidth]{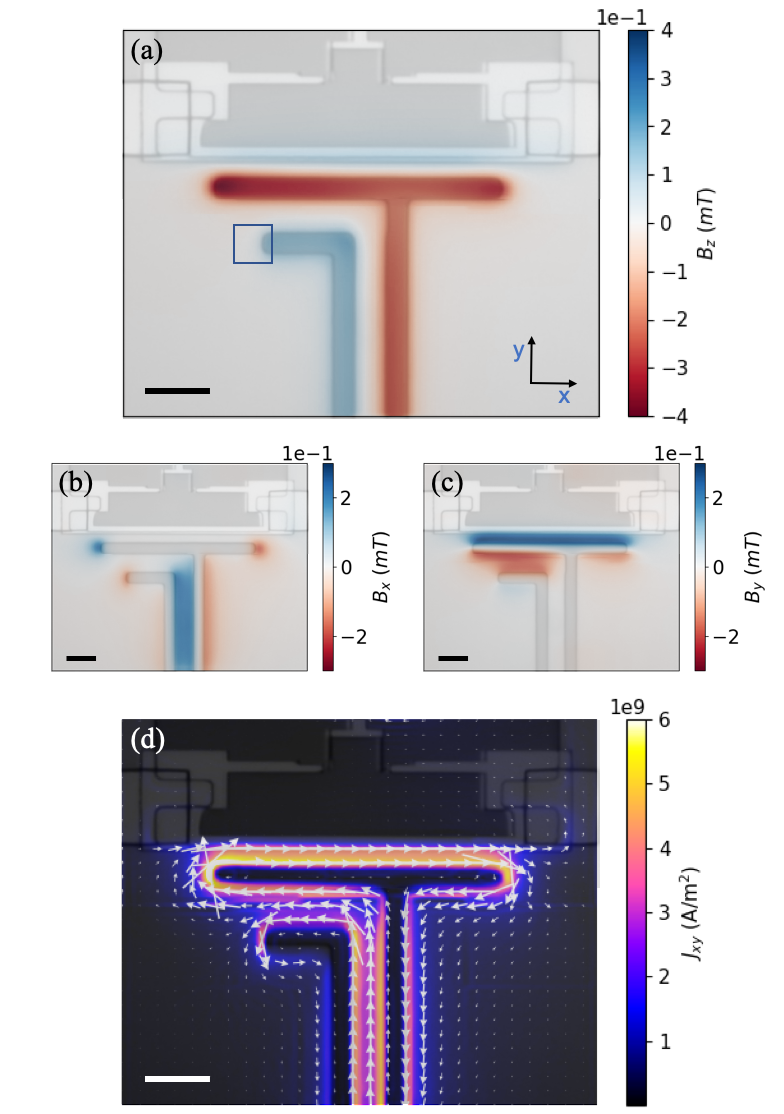}
  \caption{Magnetic field and current density maps.  (a) AC $B_z(x,y)$
    measured by the SSM at a tip-sample spacing of
    \SI{600}{\nano\meter} and generated by a current of
    \SI{3}{\milli\ampere} at \SI{9.8}{\kilo\hertz}.  All color plots
    are overlaid on an optical micrograph of the corresponding region
    of the qubit device.  The blue square denotes the scanning area
    shown in Fig.~\ref{Fig3}.  (b) $B_x(x,y)$ and (c) $B_y(x,y)$
    extracted from the measured $B_z(x,y)$.  (d) Two-dimensional
    current density $\mathbf{J}_{xy}(x,y)$ reconstructed from the
    measured $B_z(x,y)$.  The vector field represents
    $\mathbf{J}_{xy}(x,y)$, while the color scale shows its
    magnitude. Scale bars: \SI{10}{\micro\meter}.}
  \label{Fig2}
\end{figure}

Fig.~\ref{Fig2}~(a) shows a map of AC $B_z (x, y)$, produced by
\SI{3}{\milli\ampere} of current at \SI{9.8}{\kilo\hertz} applied to the
flux-control line and measured \SI{600}{\nano\meter} above the qubit
circuit.  The out-of-plane field vanishes above the superconducting
Nb, due to the Meissner effect, and is concentrated in the gaps of
exposed dielectric around the superconducting control line.
Out-of-plane field is also measured in the bottom part of the transmon
qubit's SQUID loop, as well as over the Al film defining the bottom of
the SQUID loop.  At the measurement temperature of \SI{4.2}{\kelvin},
unlike during normal operation of the qubit at \SI{20}{\milli\kelvin},
Al is not superconducting and therefore does not expel magnetic
field.  The in-plane components $B_x (x, y)$ and $B_y (x, y)$, shown
in Fig.~\ref{Fig2}~(b) and (c), can be calculated from the measured
$B_z (x, y)$ in the source-free region above the circuit, for slowly
varying magnetic fields (in the magnetostatic
limit)~\cite{lima_obtaining_2009}.

The reconstructed $\mathbf{J}_{xy} (x,y)$ is plotted as a vector field
in Fig.~\ref{Fig2}~(d), showing the flow of current out of the flux
control line and into the ground plane.  The flow pattern reveals how
current density is concentrated along the edges of the Nb control line
structures, as expected for a superconductor.  As can be seen in the
map of $B_z(x,y)$ in Fig.~\ref{Fig2}~(a), the magnetic flux coupling
into the SQUID loop of the transmon qubit, appearing as light blue
contrast in the upper part, results principally
from the return current flowing in the $x$-direction, along the top
edge, where the ground plane ends and the qubit structure begins. In
particular, the narrowness of this superconducting channel, which
provides the most direct connection between the two sides of the
circuit's ground plane, appears to concentrate the return current
density in this region, enhancing the flux produced through the
transmon structure.  By integrating the current density, we find that
\SI{2.5}{\milli\ampere} of the \SI{3.0}{\milli\ampere} flowing out of
the control line flow through this narrow channel.

The coupling strength between the control line and the transmon can be
quantified by a mutual inductance $M = \Phi_{\text{qubit}}/I$, where
$\Phi_{\text{qubit}}$ is the magnetic flux threading through the
qubit's SQUID loop and $I$ is the current flowing through the control
line.  By integrating the magnetic flux measured by our SSM in the
region directly above the qubit's SQUID, we estimate
$M = 1.1 \pm 0.2$~$\Phi_0/\text{mA}$.  This value is significantly
larger than $M = 0.62 \pm 0.1$~$\Phi_0/\text{mA}$ measured on another
identical chip at \SI{20}{\milli\kelvin} via its flux periodicity as a
function of DC current applied to the control line.  This discrepancy
is at least in part due to the difference in temperature between the
two measurements, since none of the Al circuit elements, including the
transmon SQUID loop itself or the airbridges, are superconducting at
\SI{4.2}{\kelvin}.  As a result, return currents have different
available superconducting paths along which to flow.  Another
possibile source of the discrepancy is the different configuration of
the bonds at the edges of the chip connecting the ground plane to the
system ground.  The sample measured via SSM has fewer connections that
are not uniformly distributed around the chip.  Finally, although
other SSM measurements were carried out in $B_a = 0$ after zero-field
cooling, the measurement shown in Fig.~\ref{Fig2} was carried out in
an applied field $B_a = \SI{55}{\milli\tesla}$, because the
SQUID-on-tip used in that case lacked sensitivity near zero field.  At
such fields, flux is expected to penetrate the ground plane,
potentially affecting the flow of AC return currents.

For all of these reasons, this specific measurement should be taken as
a demonstration showing that SQUID-on-tip SSM can provide detailed map
of current flow in superconducting qubit circuits.  However, if
carried out below \SI{1}{\kelvin} and on devices grounded as in normal
operating conditions, maps of current density can give a precise
picture of the desired and undesired inductive couplings.  In
addition, the uncertainty in determining mutual inductance in this
measurement is due to both the limited accuracy in the alignment of
the flux image measured via SSM with the geometry of the qubit and to
the spread of the flux over the \SI{600}{\nano\meter} separating the
plane of the qubit and the probe.  Applying SSM probes that include
simultaneous AFM measurements of sample topography, e.g.\ using probes
developed by Wyss et al.~\cite{wyss_magnetic_2022} or done with
smaller tip-sample separations would improve the measurement's
accuracy.  Ultimately, SSM done as shown here can assist in the design
of qubit circuits with increased mutual inductance.  Increasing mutual
inductance allows for the reduction in the size of the qubit SQUID
loop and, in turn, a reduction in the flux noise due to ambient
magnetic fields, making the qubit more robust.  Given the ability of
SQUID-on-tip sensors to detect a few tens of
$\text{nA}/\sqrt{\text{Hz}}$ of
current~\cite{vasyukov_scanning_2013,marchiori_nanoscale_2022},
similar measurements could also identify current paths that lead to
spurious coupling between a control line and other qubits on the chip.
Identifying the sources of such cross-talk could aid in the design of
circuits minimizing these effects.

\begin{figure}[t]
  \includegraphics[width=0.48\textwidth]{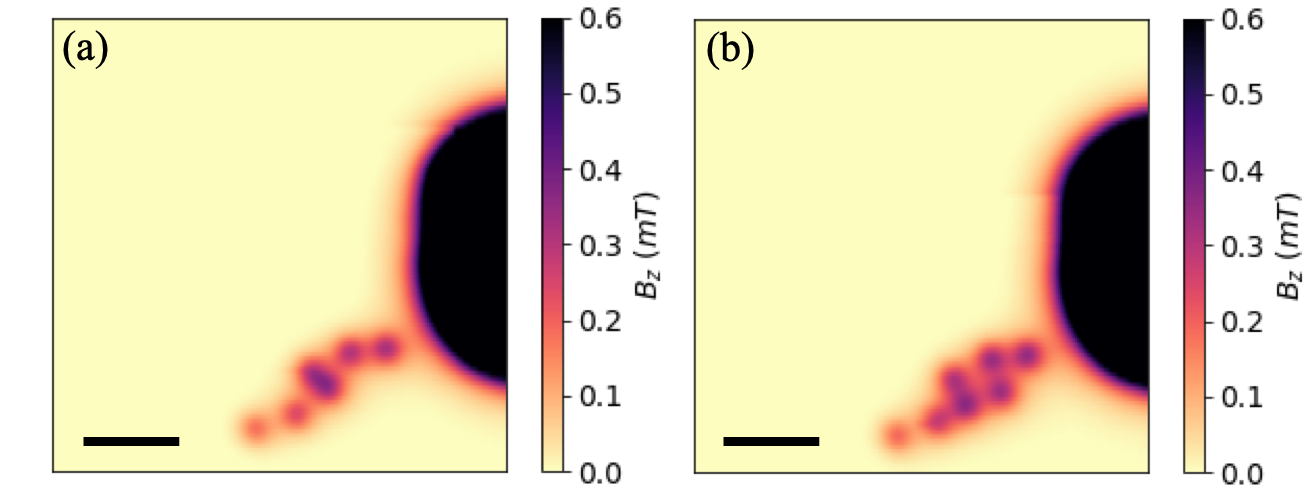}
  \caption{Location of trapped magnetic flux.  (a) A map of DC
    $B_z(x,y)$ measured after zero-field cooling followed by the
    application of an out-of-plane applied field
    $B_a= \SI{0.6}{\milli\tesla}$.  The scanned region corresponds to
    the blue box shown in Fig.~\ref{Fig2}~(a).  Dark regions show
    where flux penetrates the circuit, while light regions are
    shielded by the Meissner effect.  The dots correspond to
    superconducting vortices. (b) $B_z(x,y)$ measured with an
    additional \SI{2}{\milli\ampere} of DC current applied to the flux
    control line.  The current has shifted the position of one of the
    vortices and caused the nucleation of an additional vortex. Scale
    bars: \SI{1}{\micro\meter}.}
  \label{Fig3}
\end{figure}

In addition to studying the flow of low-frequency current out of the
qubit control lines, we also identify the locations where magnetic
flux can be trapped in the circuit, when an external magnetic field is
applied or a large DC current is sourced through the control lines.
Fig.~\ref{Fig3}~(a) shows an image of the DC $B_z (x,y)$
\SI{600}{\nano\meter} above the circuit after zero-field cooling
followed by the application of an out-of-plane field
$B_a = \SI{0.6}{\milli\tesla}$.  Dark regions represent regions of
high magnetic field, where flux penetrates, while light regions
represent vanishing magnetic field, due to the Meissner effect of the
superconducting Nb film.  $B_a = \SI{0.6}{\milli\tesla}$ corresponds
to the lowest applied field, at which flux penetrates in the form of
superconducting vortices, as seen in the lower half of
Fig.~\ref{Fig3}~(a).  In the investigated region, vortices are always
seen to penetrate near the rounded edge of the flux control line.
Both circuit geometry and defects in the Nb film are likely to play a
role in concentrating magnetic flux and determining the location of
initial vortex entry.

The addition of a DC current of \SI{2}{\milli\ampere} is then seen to
shift the position of the vortices and to introduce an additional
vortex, as shown in Fig.~\ref{Fig3}~(b). Also, two vortices which were
initially closer than our spatial resolution are seen to separate
slightly. The new vortex penetrates just to the right of the newly
separated vortices.  The fact that a DC current through the control
line, which is similar in magnitude to that used during qubit
operation, results in the perturbation of vortices and the nucleation
of a new vortex supports the possibility that trapped flux may be
responsible for some of the low-frequency noise observed in similar
qubit devices~\cite{nsanzineza_trapping_2014}.

Unstable vortex configurations could produce fluctuations in magnetic
flux and therefore in qubit frequency, ultimately leading to qubit
decoherence.  In fact, the circuit under investigation includes
flux-trapping holes, appearing as regularly spaced features in
Fig.~\ref{Fig1}~(a), except in the region around the qubit, in order
not to affect its flux controllability.  Although under normal
operation, such a superconducting circuit would be shielded from
external applied magnetic fields, vortices could penetrate as a result
of the magnetic flux produced by residual background fields due to
slightly magnetic sample-mounting components.  In addtion, the control
currents, which flow through the flux control lines and are on the
order of mA, may also result in the penetration of flux.  On the other
hand, we did not observe the entry of vortices in the same region at
$B_a = 0$ with DC currents of up to \SI{4}{\milli\ampere}.  Recent
studies have also suggested that qubit instability is the result of
two-level system fluctuators, rather than vortex
instability~\cite{klimov_fluctuations_2018,schlor_correlating_2019,burnett_decoherence_2019,muller_towards_2019}.

These experiments show the potential of SSM with a nanometer-scale
probe to reveal the details of current flow and flux control-line
coupling in superconducting qubit circuits.  In the future, comparison
of current density maps measured below $\SI{1}{\kelvin}$ with
simulations could help improve circuit design, both to optimize
couplings and to reduce unwanted cross-talk and interference between
qubits.  Extending the bandwidth of nanometer-scale SSM probes from
the kHz frequency range, which is currently possible, to hundreds of
MHz or GHz would be useful for visualizing currents and fields from
qubit control pulses and mapping the high-frequency behavior of the
circuit.  The identification of vortex entry as a result of flux
produced by large control currents also may explain instabilities
observed in qubit devices.  The identification of threshold control
currents for vortex entry may help mitigate these effects.

\begin{acknowledgements}
  We thank Sascha Martin and his team in the machine workshop of the
  Physics Department at the University of Basel.  We acknowledge the
  support of the Canton Aargau and the Swiss National Science
  Foundation via Project Grant No. 200020-159893, Sinergia Grant
  Nanoskyrmionics (Grant No. CRSII5-171003), and via a Director's
  Reserve grant of the National Centre for Competence in Research
  Quantum Science and Technology (QSIT).  S.K. acknowledges financial
  support from Fondation Jean-Jacques et Félicia Lopez-Loreta and the
  ETH Zurich Foundation.
\end{acknowledgements}

The data that support the findings of this study are available from
the corresponding author upon reasonable request.

%

\end{document}